\documentclass{article} % For LaTeX2e
\usepackage{iclr2026_conference,times}

\usepackage{amsmath,amsfonts,bm}

\def\eqref#1{equation~\ref{#1}}
\def\1{\bm{1}}

\DeclareMathAlphabet{\mathsfit}{\encodingdefault}{\sfdefault}{m}{sl}
\SetMathAlphabet{\mathsfit}{bold}{\encodingdefault}{\sfdefault}{bx}{n}

\usepackage[hidelinks]{hyperref}
\usepackage{url}
\usepackage[dvipsnames]{xcolor}
\usepackage{graphicx}
\usepackage{booktabs}

\usepackage{algorithm}     
\usepackage{algorithmic}

\newcommand{\angstrom}{\ensuremath{\text{\AA}}}

\title{Protein generation with embedding learning for motif diversification}

\author{Kevin Michalewicz$^{1,2}$\thanks{For correspondence: k.michalewicz22@imperial.ac.uk} \And Chen Jin$^{1}$ \And Philip Teare$^{1}$ \And Tom Diethe$^{1}$ \And Mauricio Barahona$^{2}$ \AND Barbara Bravi$^{2}$ \And Asher Mullokandov$^{1}$\thanks{For correspondence: asher.mullokandov@astrazeneca.com}\hspace{8cm} \\
\\ 
\hspace{-2.85cm}$^{1}$Centre for AI, Data Science \& Artificial Intelligence, Biopharma R\&D, AstraZeneca, UK\\
\hspace{-2.85cm}$^{2}$Department of Mathematics, Imperial College London, UK
}

\iclrfinalcopy % Uncomment for camera-ready version, but NOT for submission.
\begin{document}

\maketitle

\begin{abstract}
A fundamental challenge in protein design is the trade-off between generating structural diversity while preserving motif biological function. Current state-of-the-art methods, such as partial diffusion in RFdiffusion, often fail to resolve this trade-off: small perturbations yield motifs nearly identical to the native structure, whereas larger perturbations violate the geometric constraints necessary for biological function. We introduce Protein Generation with Embedding Learning (PGEL), a general framework that learns high-dimensional embeddings encoding sequence and structural features of a target motif in the representation space of a diffusion model's frozen denoiser, and then enhances motif diversity by introducing controlled perturbations in the embedding space. PGEL is thus able to loosen geometric constraints while satisfying typical design metrics, leading to more diverse yet viable structures. We demonstrate PGEL on three representative cases: a monomer, a protein-protein interface, and a cancer-related transcription factor complex. In all cases, PGEL achieves greater structural diversity, better designability, and improved self-consistency, as compared to partial diffusion. Our results establish PGEL as a general strategy for embedding-driven protein generation allowing for systematic, viable diversification of functional motifs.

%A central challenge in protein design is balancing structural diversity with preservation of functional motifs. In RFdiffusion, diversity is typically introduced by applying random perturbations in the seed space, but this approach is limited: small perturbations yield near-native motifs, while large ones break functional geometry. Moreover, retraining or fine-tuning diffusion models on diverse datasets is computationally costly and often infeasible.
%
%We present Protein Generation with Embedding Learning (PGEL), a framework that addresses these limitations by learning high-dimensional embeddings of target motifs in the representation space of a frozen diffusion denoiser. PGEL enhances diversity by introducing controlled perturbations in this embedding space, which is richer and more expressive than seed space, while keeping the base model fixed. This design yields a cost-efficient solution that loosens geometric constraints without sacrificing functional validity, thereby enabling systematic and scalable motif diversification.

% We evaluate PGEL on three representative cases: a monomer, a protein–protein interface, and a cancer-related transcription factor complex. Across all settings, PGEL achieves greater structural diversity, improved designability, and better self-consistency compared to partial diffusion. These results highlight embedding learning as a practical strategy for exploiting the underexplored design space of large pretrained protein models, offering a general and efficient path toward functional protein diversification.
\end{abstract}

\section{Introduction}

Designing proteins that achieve precise biological functions while allowing for structural diversity has long been a central goal in computational protein design. Recent advances in structure prediction models like AlphaFold~\citep{Jumper2021,Abramson2024}, RoseTTAFold~\citep{Baek2021}, ESMFold~\citep{LinESM2023}
%, OmegaFold~\citep{Wu2022}
and Boltz~\citep{Wohlwend2024,Passaro2025} have revolutionized protein generative models and paved the way for improved diffusion models in protein design. Among them, RFdiffusion~\citep{Watson2023}, which results from fine-tuning RoseTTAFold, has shown strong performance in both unconditional and conditional  generation. 

Yet, targeted local modification remains a challenge. A common approach is \emph{partial diffusion} in RFdiffusion, in which a native or designed structure undergoes only a few denoising steps to induce diversification~\citep{Watson2023,VazquezTorres2024}. However, this method faces a fundamental diversity-fidelity trade-off: small structural perturbations keep near-native conformations, but lack diversity, while larger perturbations induce excessive geometric drift that disrupts functional features~\citep{Lin2024}. Overcoming this limitation requires rethinking how diffusion models can introduce controlled variation while still anchoring designs to essential geometric constraints.

A promising direction comes from recent advances in conditional image generation. Models such as Stable Diffusion and Latent Diffusion Models (LDMs) generate images from noise guided by text prompts~\citep{ho2020denoising,Rombach2022}. Beyond standard prompting, textual inversion learns new prompt embeddings to represent unseen visual concepts~\citep{Gal2022,Jin2024}. Once learned, these embeddings can be diversified to generate outputs that preserve the original concept while exploring novel variations. Here we adopt this embedding-centric view in the context of protein generation.

We present Protein Generation with Embedding Learning (PGEL), a general framework representing the first adaptation of textual inversion principles to protein diffusion models. PGEL introduces two key approaches with broad applicability: (1) learning high-dimensional embeddings that capture the sequence and structural characteristics of target protein regions of interest, thus shifting the paradigm from coordinate-space to embedding-space perturbations, and (2) relaxing evolutionary and structural constraints by masking  embeddings. Although we present our work here using RFdiffusion's representation space, our method is general and readily adaptable to other protein diffusion models, and can thus leverage the rich representational capacity of pre-trained diffusion models without expensive retraining or fine-tuning. 

We focus on \textit{motif diversification}. Here, a \textit{motif} denotes a set of residues with a particular geometric arrangement, which may govern functional activity. The objective is to generate a set of backbones that keep a fixed scaffold in real space within tight bounds, and realize diverse yet functionally plausible conformations of a motif, while satisfying standard designability criteria so that downstream sequence design and structure prediction can recover the intended structures. 
Some existing approaches address related challenges, but differ in scope and implementation: structure inpainting methods (\textit{e.g.}, masked region generation) fully marginalize a region by masking and regenerating it \textit{de novo}, discarding the specific native geometry~\citep{zhang2023framedipt}, whereas flexible backbone loop remodeling in Rosetta (KIC/Next-Generation KIC) samples local conformations under explicit geometric and energetic restraints to achieve high-fidelity but relatively localized exploration~\citep{Mandell2009,Stein2013,leman2020macromolecular}. %; (c) hydrogen-bond-network or "buttressed" loop stabilization strategies rigidify or pre-filter loop geometries via designed interaction networks to retain stability during diversification~\citep{Correia2014}.
Hence these tools do not explicitly target controlled exploration of a \emph{neighborhood} around an existing functional motif while keeping a surrounding scaffold nearly fixed.

Thus, we compare chiefly to partial diffusion in RFdiffusion, the prevailing stochastic baseline for local variation which has been recently applied in therapeutically relevant design settings, including \textit{de novo} creation of high-affinity peptide binders and venom toxin neutralizers~\citep{VazquezTorres2024,vazquez2025novo}. Across three representative scenarios involving a monomeric protein (calmodulin), a protein–protein binding site (barstar-barnase), and a p53 binder within the p53-MDM2 complex, PGEL (1000 samples) produces more designable structures (motif pLDDT $\geq 70$, scRMSD $\leq 1\angstrom$, mRMSD $\leq 2\angstrom$) than partial diffusion: 1000 \textit{vs} 411 (monomer), 990 \textit{vs} 331 (binding site), and 802 \textit{vs} 252 (binder). PGEL also yields more structurally diverse TM-score clusters distinguishable from native, and shows better self-consistency after inverse folding and refolding (meeting mRMSD and pAE thresholds), while maintaining predicted binding affinities comparable to native and exceeding those obtained with partial diffusion. Our results support embedding learning combined with masking as a general, efficient strategy for systematic motif diversification.

\section{Background}

%\textbf{Protein design.} Structure prediction models like AlphaFold~\citep{Jumper2021,Abramson2024}, RoseTTAFold~\citep{Baek2021}, ESMFold~\citep{LinESM2023}, OmegaFold~\citep{Wu2022} and Boltz~\citep{Wohlwend2024,Passaro2025} have revolutionized protein generative models and paved the way for improved diffusion models in protein design. For instance, RFdiffusion~\citep{Watson2023} results from fine-tuning RoseTTAFold with remarkable success in both unconditional and conditional protein generation. Similarly, FrameDiff~\citep{Yim2023} introduces an SE(3)-invariant diffusion model, integrating a structural module based on AlphaFold2, and delivers competitive results with those of RFdiffusion. Genie~\citep{LinGenie2023} and FoldingDiff~\citep{Wu2024} represent proteins as cloud of oriented reference frames or using internal angles, respectively, and have successfully trained diffusion models to generate designable and diverse protein backbones.

\textbf{Functional motifs.} 
Conditional generation around functional residues, often framed as \textit{motif scaffolding}, has been a focal point for recent protein design methods. In that setting, the motif and scaffold are defined as disjoint subsets with the scaffold varied while the motif geometry is preserved. Approaches like RFdiffusion (where the motif coordinates are fixed), the Monte Carlo-based Twisted Diffusion Sampler~\citep{Wu2023} applied to FrameDiff, and Genie2~\citep{Lin2024} have made progress on this task, though performance remains task-dependent and can yield few or no backbones meeting success criteria in specific cases. In our motif diversification task, the scaffold is held fixed and the motif is diversified to explore multiple, function-preserving geometric realizations, enabling improvements in \textit{e.g.}, affinity, specificity or stability, while maintaining the broader structural context.
% A common application of protein design, particularly for therapeutic applications, is conditional generation focused on a set of residues encoding a specific function (\textit{motif scaffolding} task). This involves defining two disjoint subsets of protein residues: the motif and the scaffold. The goal is to produce scaffold diversity while maintaining motif function and physical feasibility. RFdiffusion fixes the motif coordinates and reverses the noising process of the rest of the structure. The Monte Carlo-based Twisted Diffusion Sampler~\citep{Wu2023} was applied to FrameDiff, obtaining a comparable success rate to RFdiffusion. Meanwhile, Genie2~\citep{Lin2024} has upgraded its training capabilities to solve motif scaffolding tasks and improves diversity relative to its predecessors, though performance is highly task-dependent and, in some cases, yields no backbones that meet success criteria. In our proposed motif diversification task, the scaffold is held fixed and the motif is diversified to explore multiple, function-preserving geometric realizations, enabling improvements in affinity, specificity, or stability while maintaining the broader structural context.

\textbf{Protein embeddings.} The limited availability of structural data motivated the development of models that transform sequences into sequence embeddings that encode structural information. These embeddings have been employed for various tasks such as property prediction using a Gaussian Process regression model~\citep{Yang2018} and residue-residue contact prediction via a Bidirectional Long Short-Term Memory (BiLSTM) architecture~\citep{Bepler2019}. Transformers~\citep{Vaswani2017} have been used in generating sequence embeddings, including for antibody-specific applications like paratope prediction~\citep{Leem2022}. Transfer learning has also been shown to significantly improve performance across architectures by enabling the use of pre-trained embeddings that capture fundamental sequence-structure relationships~\citep{Detlefsen2022}.
Other approaches explicitly include structural information~\citep{Ali2024}, such as contact maps-derived embeddings, and have shown enhanced performance in particular downstream tasks such as structure similarity assessment~\citep{Kandathil2025}, structure searching~\citep{Greener2024}, property prediction~\citep{Blaabjerg2024,Danner2025}, and domain classification~\citep{Lau2024}. Similarly, protein function annotation and local flexibility prediction have benefited from Graph Convolutional Networks, which combine structure-derived graphs to propagate contextual signals from protein sequence embeddings obtained with pre-trained models~\citep{Gligorijevic2021,Michalewicz2025}.

\textbf{Diffusion models for proteins.} %The first successful application of Denoising Diffusion Probabilistic Models (DDPMs) in protein design was  RFdiffusion~\citep{Watson2023,Bennett2024}.
Earlier works adapted Denoising Diffusion Probabilistic Models (DDPMs) to protein design by conditioning on local structural elements or coarse fold constraints~\citep{wu2024protein,anand2022protein,Trippe2023,luo2022antigen} yet, while encouraging, they produced few sequences that refolded to target backbones. RFdiffusion subsequently emerged as the diffusion approach that reliably yields designable structures and sequences that recover the intended geometry. In RFdiffusion, a highly accurate protein structure prediction method (RoseTTAFold~\citep{Baek2021}) is fine-tuned to undo random perturbations of atomic coordinates introduced via 3D Gaussian noise (\textit{i.e.}, to denoise). 
RFdiffusion can be constrained to specific binding targets, or symmetry specifications, and once trained it can be viewed as a \textit{frozen denoiser}. RoseTTAFold/AlphaFold-style models (including RFdiffusion) learn so-called \textit{state} and \textit{pair} embeddings (related to per-residue and residue-residue properties of the protein structure, respectively) and MSA embeddings related to multiple sequence alignment~\citep{Jumper2021}.

\textbf{Textual inversion.} \cite{Gal2022} builds on LDMs~\citep{Rombach2022}, a specific class of DDPMs, to perform textual inversion. In the context of text-to-image models, let $x$ represent an image, $s$ a text prompt, $\epsilon_{\theta}$ a pre-trained denoising network, and $\varepsilon$ an image encoder. LDMs aim to minimize the following loss:
\begin{equation}
\label{eq:LDM_loss}
\mathcal{L}_{\mathrm{LDM}}:=\mathbb{E}_{z\sim\varepsilon(x), s, \epsilon\sim\mathcal{N}(0,1), t}\left[\lVert\epsilon-\epsilon_{\theta}(z_t, t, c_{\theta}(s)) \rVert_2^2\right]\end{equation}
Here, $c_{\theta}(s)$ represents a pre-trained text encoder that conditions the denoiser $\epsilon_{\theta}$ based on the text prompt $s$, and $z_t$ is a noised version of the image embedding $z$ at timestep $t$.
The goal of textual inversion is to learn a new text embedding $v_*$ corresponding to a particular concept $s_*$ such that it minimizes the LDM loss (\eqref{eq:LDM_loss}). This means conditioning $\epsilon_{\theta}$ on $v_*$ so the generated image $\tilde{x}$ closely resembles the original image $x$. Neutral prompts, such as ``A photo of $s_*$" or ``A portrait of $s_*$" are used while keeping $\epsilon_{\theta}$ and $c_{\theta}$ frozen. Multi-Concept Prompt Learning~\citep{Jin2024} extends this idea to handle multiple concepts by incorporating three regularization techniques: attention masking, bind adjective, and prompts contrastive loss.

\section{Methods}

We now present our method, \textit{Protein Generation with Embedding Learning (PGEL)}, and describe how we learn the embedding representation of a motif in Section~\ref{subsec:PGEL}.  In Section~\ref{subsec:diversity}, we propose an approach to increase motif diversity, and Section~\ref{subsec:metrics} details the evaluation metrics.

\subsection{Protein Generation with Embedding Learning (PGEL)}\label{subsec:PGEL}

We generalize the notion of textual inversion with LDMs to proteins, treating the structure as analogous to an image, and the sequence as analogous to a text prompt. Let $R_*$ be a region of interest, or \textit{motif}, defined as a continuous or discontinuous set of $L_*$ amino acids within a protein. The motif has structure $x_*$ and sequence $s_*$, where the coordinates of $x_*$ are obtained from an experimental Protein Data Bank (PDB) entry, and the sequence $s_*$ is \emph{masked} when passed as an input to PGEL, \textit{i.e.}, the amino acid range of the motif is specified, but not its exact composition. 

PGEL learns a representation of $R_*$ in embedding space, which we denote as $v_*$. The remainder of the protein constitutes the \textit{scaffold}, with structure $x_c$ and sequence $s_c$ of length $L_c$, from which the protein LDM frozen \textsc{Encoder} computes an embedding representation $v_c$. 

\begin{figure}[h]
\begin{center}
%\framebox[4.0in]{$\;$}
%\fbox{\rule[-.5cm]{0cm}{4cm} \rule[-.5cm]{4cm}{0cm}}
\includegraphics[width=\linewidth,clip,trim=0cm 23.75cm 17cm 0cm]{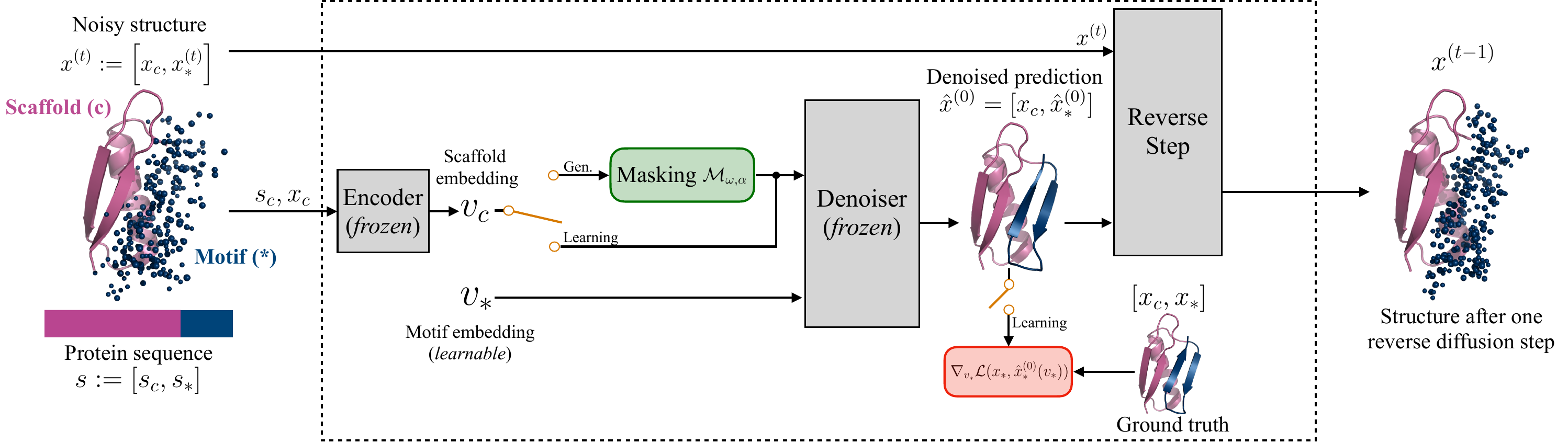}
\end{center}
\caption{Outline of the PGEL learning and generation procedures during one reverse diffusion step.}
\label{fig:1}
\end{figure}

The procedure (see Figure~\ref{fig:1} and Algorithm~\ref{alg:PGEL-ims}) starts by building a noised protein structure in which the scaffold coordinates are retained while the motif coordinates are subjected to $T$ rounds of Gaussian noise injection, following~\cite{Trippe2023}. At each timestep $t$, the protein LDM frozen \textsc{Denoiser} predicts a denoised motif structure $\hat{x}_*^{(0)}$, conditioned jointly on the learnable motif embedding $v_*$ and the fixed embedding $v_c$. These embeddings include state, pair and MSA embeddings. Then, by using structure $x^{(t)}$ and the intermediate structure $[x_c,\hat{x}_*^{(0)}]$, a reverse diffusion step \textsc{ReverseStep}, which does not contain any learnable parameters, yields $x^{(t-1)}$ (see Algorithm~\ref{alg:ReverseStep}). In practice, we employ pre-trained building blocks of RFdiffusion for both the \textsc{Encoder} and \textsc{Denoiser}, though alternative models could be substituted if desired.

\textbf{Embedding optimization.} The embedding $v_*$ is learned by minimizing:
\begin{equation}\mathcal{L} =
\mathcal{L}_{\text{MSE}} +
\lambda_{\text{DM}}  \mathcal{L}_{\text{DM}}  +
\lambda_{\text{torsion}} \mathcal{L}_{\text{torsion}}\end{equation}
%
%where $\lambda_{\text{DM}},\lambda_{\text{torsion}}\in\mathbb{R}_{\geq 0}$ are regularization parameters. 
This loss function is composed of three terms, described hereafter, which compare different features of the ground truth structure $x_*$ and the predicted structure $\hat{x}^{(0)}_*(v_*)$ of the motif with the coefficients $\lambda_{\text{DM}},\lambda_{\text{torsion}}\in\mathbb{R}_{\geq 0}$ controlling the relative weight of the terms.

\textbf{Data fidelity term (backbone atoms).} For each motif residue $i \in R_*$ we consider the $A = 4$ backbone atoms (nitrogen N, $\alpha$-carbon $\text{C}_{\alpha}$, carbon C, oxygen O). Let $\hat{x}^{(0)}_{i,a} \in \mathbb{R}^3$ denote the predicted position of atom $a$ in residue $i$ and $x_{i,a}\in\mathbb{R}^3$ its ground truth counterpart. We then compute the mean squared error (MSE)  between the backbone atoms of the ground truth and predicted motif:
\begin{equation}
\mathcal{L}_{\text{MSE}}(x_*, \hat{x}_*^{(0)}(v_*)) =
\frac{1}{AL_*} \sum_{i \in R_*} \sum_{a \in \{\text{N}, \text{C}_{\alpha}, \text{C}, \text{O}\}}
\left\| \hat{x}^{(0)}_{i,a}(v_*) - x_{i,a} \right\|^2
\end{equation} 
\textbf{Distance matrix between $\alpha$-carbons.} Let $\hat{x}^{(0)}_{i,{\text{C}_{\alpha}}}\in\mathbb{R}^3$ denote the predicted position of the $\alpha$-carbon atom in residue $i$, and $x_{i,{\text{C}_{\alpha}}}\in\mathbb{R}^3$ its ground truth counterpart. We define the following loss term based on $\alpha$-carbon Distance Matrices (DM), inspired by the distrogram notion~\citep{Senior2020}:
\begin{equation}
\mathcal{L}_{\text{DM}}(x_*, \hat{x}_*^{(0)}(v_*)) =
\frac{1}{L_*^2} \sum_{i \in R_*} \sum_{j \in R_*}
\left(
\left\| \hat{x}^{(0)}_{i,\text{C}_{\alpha}}(v_*) - \hat{x}^{(0)}_{j,\text{C}_{\alpha}}(v_*) \right\|
-
\left\| x_{i,\text{C}_{\alpha}} - x_{j,\text{C}_{\alpha}} \right\|
\right)^2 
\end{equation}
In contrast to $\mathcal{L}_{\text{MSE}}$, $\mathcal{L}_{\text{DM}}$ is invariant under rigid motions (translations and rotations), thus encouraging global shape consistency.

\textbf{Backbone torsion angles.} Let $\hat{\phi}_i$ and $\hat{\psi}_i$ denote the predicted backbone torsion angles at residue $i$, computed from $\hat{x}_*^{(0)}(v_*)$, and let $\phi_i$ and $\psi_i$ be the corresponding ground truth values~\citep{Ramachandran1963}. We impose a constraint on angular torsions through a cosine-based loss term akin to that of AlphaFold~\citep{Jumper2021}:
\begin{equation}
\mathcal{L}_{\text{torsion}}(x_*, \hat{x}_*^{(0)}(v_*)) =
\frac{1}{L_*-2} \sum_{i=2}^{L_*-1}
\left[
1 - \cos\!\left(\hat{\phi}_{i}(v_*) - \phi_{i}\right)
+ 1 - \cos\!\left(\hat{\psi}_{i}(v_*) - \psi_{i}\right)
\right ]
\end{equation}
This term penalizes sterically implausible geometries, helping improve performance under the predicted local distance difference test (pLDDT). We do not include the third backbone angle $\omega$ as it is typically considered fixed at 180 degrees~\citep{Cutello2006}.

\begin{algorithm}[!h]
   \caption{PGEL -- Embedding learning}
   \label{alg:PGEL-ims}
\begin{algorithmic}
   \STATE {\bfseries Input:} region of interest/motif $R_*$ with masked sequence $s_*$ and structure $x_*$, fixed scaffold with sequence $s_c$ and structure  $x_c$, pre-trained \textsc{Encoder} and  \textsc{Denoiser}.
   \STATE {\bfseries Output:} learned embedding $v_*$ for region $R_*$.
   \STATE initialize $v_*$ with zeros.
   \WHILE{not converged}
   \STATE Build noised structure $x^{(T)}:=[x_c,x_*^{(T)}]$ with associated sequence $s:=[s_c, s_*]$.
   \STATE $v_c=\textsc{Encoder}(s_c,x_c)$ 
   \FOR{$t=T$ {\bfseries down to} $1$}
   \STATE $\hat{x}^{(0)}_*=\mathrm{\textsc{Denoiser}}(v_c, v_*)$
   \STATE $x^{(t-1)}=\textsc{ReverseStep}(x^{(t)}, [x_c,\hat{x}_*^{(0)}])$
   \STATE Update $v_*$ by taking a gradient step 
   %\STATE 
   $\nabla_{v_*}\mathcal{L}(x_*,\hat{x}_*^{(0)}(v_*))$
   \ENDFOR
   \ENDWHILE
   \STATE \textbf{Return} $v_*$
\end{algorithmic}
\end{algorithm}

Once the embeddings are learned, we employ Algorithm~\ref{alg:PGEL-gen} to generate novel proteins containing a diversification of the region of interest $R_*$ (see Figure~\ref{fig:1}).  

\begin{algorithm}[!h]
   \caption{PGEL -- Generation with embedding masking}
   \label{alg:PGEL-gen}
\begin{algorithmic}
   \STATE {\bfseries Input:} region of interest/motif $R_*$ with learned embeddings $v_*$, fixed scaffold with sequence $s_c$ and structure  $x_c$, pre-trained \textsc{Encoder} and  \textsc{Denoiser}.
   \STATE {\bfseries Output:} generated structure.
   \STATE Build noised structure $x^{(T)}:=[x_c,x_*^{(T)}]$. \STATE Draw at random the sample masking type $\omega\sim \mathrm{Ber}(\frac{1}2{})$ (row if $0$, column if $1$).
   \STATE Sample masking rate $\alpha\sim\mathcal{U}[0,1]$.
   \STATE Define $\mathcal{M_{\omega,\alpha}(\cdot)}$ as a zero mask with type $\omega$ and rate $\alpha$.
   \STATE $v_c=\textsc{Encoder}(s_c,x_c)$
   \FOR{$t=T$ {\bfseries down to} $1$}
   \STATE $\hat{x}^{(0)}_*=\textsc{Denoiser}(\mathcal{M}_{\omega,\alpha}(v_c), v_*)$
   \STATE $x^{(t-1)}=\textsc{ReverseStep}(x^{(t)}, [x_c,\hat{x}_*^{(0)}])$
   \ENDFOR
   \STATE \textbf{Return} $x^{(0)}$
\end{algorithmic}
\end{algorithm}

\subsection{Enhancing the diversity of generated motifs}\label{subsec:diversity}

MSA embeddings in sequence-to-structure predictors contain evolutionary covariation information about residues, thereby capturing geometric constraints such as residue proximity. In RFdiffusion, however, such embeddings are derived solely from the input sequence $s:=[s_c, s_*]$ rather than from a full stack of aligned sequences, and can be represented as a $d_{\mathrm{MSA}}\times L$ matrix, where $d_{\mathrm{MSA}}$ is the depth of the MSA embeddings and $L:=L_*+L_c$ the total protein length. With PGEL, we show that the diversity of generated structures can be increased by applying perturbations to the scaffold MSA embeddings $v_c\in\mathbb{R}^{d_{\mathrm{MSA}}\times L_c}$. These embeddings couple through attention mechanisms with state and pair embeddings produced by an internal RFdiffusion encoder, and also interact with the learned motif embedding $v_*$, which provides an independent conditioning signal for the frozen denoiser. 

%[incorporate something like this?: "we learn a motif embedding v* that conditions the denoiser to elicit the desired motif geometry (across the reverse steps). This is analogous to learning a “concept embedding” for the motif."]

%%%  maybe you could cut out this bit...[
%rather than learned (see~\cite{Watson2023}).
%% ] up to here

\textbf{Embedding masking.} We studied the effect of applying  zero masks, \textit{i.e.} masks zeroing specific elements, to the scaffold MSA embeddings during generation (see Algorithm~\ref{alg:PGEL-gen}). \textit{Row masking} corresponds to masking specific features for all residues, whereas \textit{column masking} zeroes out all features of specific residues. Both strategies lift some constraints on inter-residue distances, and modulate which co-variation patterns remain accessible. We sample $\omega\sim\mathrm{Ber}(\frac{1}{2})$ to choose the masking mode ($\omega=0$ for row masking and $\omega=1$ for column masking) and $\alpha\sim\mathcal{U}[0,1]$, the masking rate, to set the fraction of rows or columns masked. This defines the operator $\mathcal{M}_{\omega,\alpha}(\cdot)$, which implements zero masking with type $\omega$ and rate $\alpha$. As such, masking $v_c$ relaxes the geometric constraints of the generated motif.

\subsection{Evaluation metrics}\label{subsec:metrics}

\begin{figure}[h]
\begin{center}
%\framebox[4.0in]{$\;$}
%\fbox{\rule[-.5cm]{0cm}{4cm} \rule[-.5cm]{4cm}{0cm}}
\includegraphics[width=\linewidth,clip,trim=0cm 25cm 0cm 0cm]{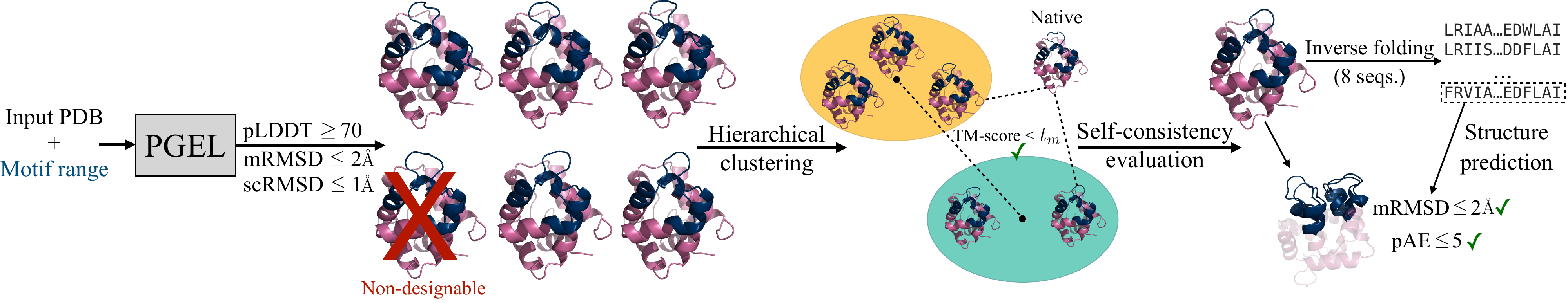}
\end{center}
\caption{Summary of the evaluation metrics. PGEL takes as input a PDB entry and the amino acid range corresponding to the motif. From 1000 PGEL-generated backbones, designable candidates are filtered by root mean square deviation (RMSD) and pLDDT thresholds, and structural diversity is assessed via hierarchical clustering. Backbones are also required to be \textit{distinguishable} from the native. Cluster representatives undergo self-consistency evaluation: sequences assigned to the designable backbones with ProteinMPNN are refolded, and at least one predicted structure must satisfy set mRMSD and predicted alignment error (pAE) conditions relative to the generated backbone.}
\label{fig:2}
\end{figure}

\textbf{Designability.} To quantify designability in the motif diversification task, we first require a motif $\mathrm{pLDDT} \geq 70$ as computed by an RFdiffusion internal block, following the threshold adopted for related tasks by~\cite{Lin2024}. We also require the scaffold RMSD to be $\mathrm{scRMSD} \leq 1\angstrom$, to ensure that the residues surrounding the motif remain fixed. Finally, we set the threshold for the motif RMSD to $\mathrm{mRMSD} \leq 2\angstrom$, to allow for structural diversification of the motif. 
%while preventing the generation of backbone conformations unrelated to the native's function.

\textbf{Diversity.} To quantify structural diversity among generated proteins (Figure~\ref{fig:2}), we compute the pairwise TM-scores~\citep{Zhang2004} across all designable candidates, and employ hierarchical clustering~\citep{Lin2024} with several linkage thresholds $t_m$ to group similar backbones under this score. Diversity is measured by the number of clusters. We evaluate also the TM-score with respect to the native motif: a cluster is considered \textit{distinguishable} from native if, for TM-score threshold $t_m\in[0,1]$, at least one cluster member exhibits lower similarity than $t_m$ relative to the native. This analysis ensures that we capture 
%not only the capacity to assign sequences but also 
the structural distinctiveness of the backbones.

\textbf{Self-consistency.} 
%For a selected TM-score threshold $t_m$, which may vary depending on the application, we consider 
For the distinguishable backbones we use the procedure in~\cite{Trippe2023} based on inverse folding to assess self-consistency between generated and predicted structures. Specifically, we use ProteinMPNN with default parameters~\citep{Dauparas2022} to assign 8 plausible sequences to each backbone, followed by a sequence-to-structure model, here AlphaFold3~\citep{Abramson2024}, to predict 8 full proteins. A designed backbone is deemed self-consistent if it satisfies for at least one of the 8 predicted structures: mRMSD $\leq 2\angstrom$ and pAE $\leq 5$ (Figure~\ref{fig:2}). Previous studies included this procedure under the designability assessment~\citep{Watson2023,Lin2024}. However, this  is computationally expensive and, when prioritizing diversity, often inefficient: many backbones either fail the initial scRMSD or mRMSD filters or exhibit negligible structural diversity. In motif diversification, diversity among generated proteins and distinguishability with respect to the native are decisive. We therefore invert the pipeline to enforce these criteria first and reserve the costly self-consistency evaluation only for diverse candidates.

\textbf{Binding affinity.} For protein-protein complexes, we run \textsc{PRODIGY}~\citep{Vangone2015,Xue2016} to estimate the binding affinity $\Delta G$ expressed in $\text{kcal}/\text{mol}$, with larger $\lvert\Delta G\rvert$ values indicating stronger binding. It is desirable that new designs present binding affinity values comparable to, or larger in magnitude than, those of the native complex. Note that learning is not optimized to enhance binding affinity; rather, this serves as an \textit{a posteriori} assessment.

\section{Experiments}

We focus on three representative test cases proposed in~\cite{Watson2023} for different tasks: (1)~Calmodulin, a monomer that plays a pivotal role in regulating the activity of nearly 100 diverse target enzymes and structural proteins~\citep{Fallon2003}; (2)~the barstar-barnase complex, in which the binding interface of barstar was diversified to probe its interaction with the extracellular ribonuclease barnase~\citep{Caro2023}; (3)~the cancer-related transcription factor p53 bound to its negative regulator MDM2~\citep{Klein2004,Li2010}.

\subsection{Protocol}

We established a protocol to systematically compare our method with RFdiffusion's partial diffusion using the metrics introduced in Section~\ref{subsec:metrics}. For partial diffusion, we generated 1000 protein backbones by uniformly sampling the number of diffusion timesteps, $T\sim\mathcal{U}\lbrace 2,3,\dots,49\rbrace$, as $T=50$ corresponds to the full diffusion process in RFdiffusion. In this way, we cover a spectrum of structural perturbations ranging from near-native backbones to unrelated ones. For PGEL, we performed the learning of $v_*$ with Stochastic Gradient Descent with learning rate $l_r=4\times10^{-4}$ and momentum $p=0.9$, $\lambda_{\mathrm{DM}}=0.01\text{ and }\lambda_{\mathrm{torsion}}=0.05$ (Algorithm~\ref{alg:PGEL-ims}), and we then generated 1000 protein backbones (Algorithm~\ref{alg:PGEL-gen}).

For both sets of 1000 generated structures, we evaluated designability and, among those deemed designable, we computed TM-scores between all generated motifs and with respect to the native structure. We then plotted the number of clusters as a function of the TM-score. For structures that were designable and diverse according to a typical TM-score threshold $t_m=0.6$~\citep{Lin2024}, we performed inverse folding through ProteinMPNN to generate compatible sequences, followed by AlphaFold3 inference to assess whether the predicted sequences refolded into the intended backbones, fulfilling the self-consistency requirement defined in Section~\ref{subsec:metrics}.

\begin{table}[t]
\caption{Comparison of partial diffusion in RFdiffusion and PGEL. The number of self-consistent clusters (diversity) is computed at TM-score threshold $t_m=0.6$.}
\label{methods-table}
\begin{center}
\begin{tabular}{l|ccc|ccc|}
\cline{2-7}
\multicolumn{1}{c|}{}&%{\bf Method} &
\multicolumn{3}{c|}{\bf Designability} &
\multicolumn{3}{c|}{\bf Diversity } \\
\multicolumn{1}{c|}{} &
\multicolumn{3}{c|}{\small{(No. of  viable structures out of 1000)}} &
\multicolumn{3}{c|}{\small{(No. of self-consistent clusters)} }\\
\cline{2-7}
 & Monomer & Binding site & Binder
 & Monomer & Binding site & Binder \\
\hline
Partial diffusion & 411 & 331 & 252 & 0  & 6 & 1 \\
PGEL              & 1000 & 990 & 802 &  2 & 10 & 6 \\
\hline
\end{tabular}
\end{center}
\end{table}

\subsection{Example 1: monomer}

Calmodulin (PDB entry: 1PRW), a monomeric protein containing a double EF-hand motif spanning residues 16--35 and 52--71, was considered as the representative test case for single-chain proteins. All the 1000 backbones candidates generated by PGEL resulted to be designable, well exceeding the 411 obtained by partial diffusion (Table~\ref{methods-table}). All of the backbones generated by partial diffusion satisfied the pLDDT constraint, consistent with the fact that RFdiffusion's training favors high-confidence local structures, but 589 of them failed to meet the expected motif RMSD threshold. In these cases, the added noise during diffusion excessively perturbed the initial backbone, leading to conformations that no longer preserved the intended geometry of the EF-hand motif.

\begin{figure}[h]
\begin{center}
%\framebox[4.0in]{$\;$}
%\fbox{\rule[-.5cm]{0cm}{4cm} \rule[-.5cm]{4cm}{0cm}}
\includegraphics[width=\linewidth,clip,trim=0cm 23cm 3cm 0cm]{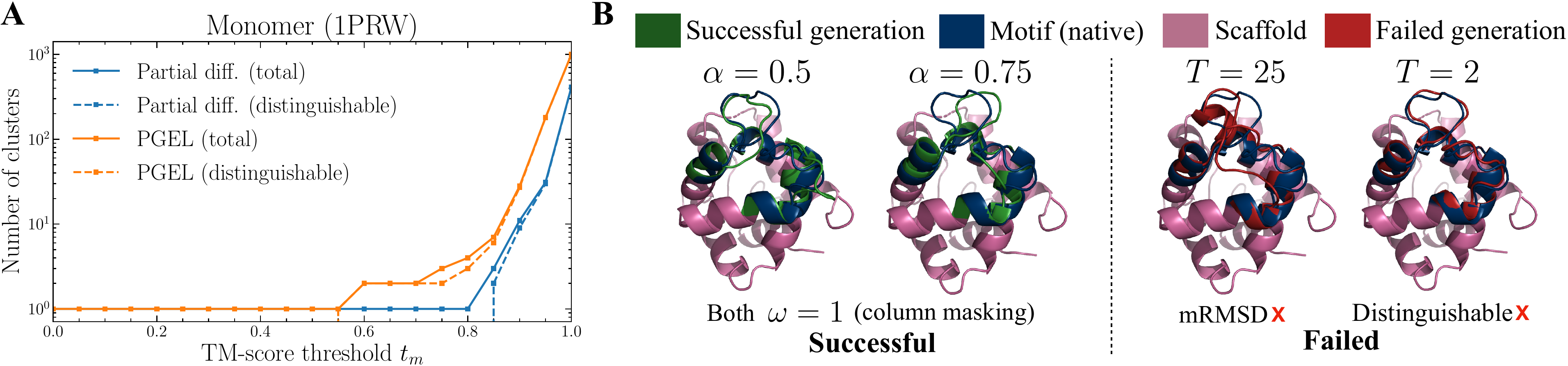}
\end{center}
\caption{Results for example 1. (A) Number of clusters, both total and distinguishable from native, as a function of the TM-score threshold $t_m$ for PGEL and partial diffusion. (B) \textit{Left:} two successful PGEL designs at $t_m=0.6$ using column masking with rates $\alpha=0.5$ and $\alpha=0.75$. \textit{Right:} two partial diffusion failed backbones at $t_m=0.6$, one obtained with $T=25$ timesteps that violates the motif RMSD constraint, and one with $T=2$ timesteps that is not distinguishable from the native.}
\label{fig:3}
\end{figure}

We then evaluated the structural diversity of the designable backbones, recording the number of clusters as a function of the TM-score threshold (Figure~\ref{fig:3}A). PGEL consistently produced a higher number of clusters across thresholds, demonstrating that embedding perturbations through masking introduce greater variability in backbone conformations. On the other hand, partial diffusion yielded structures too similar to the native backbone, and hence not distinguishable from it.

We carried out the self-consistency assessment at $t_m=0.6$, as per protocol. For PGEL, the two clusters had backbones that successfully refolded into the intended conformations after sequence design and AlphaFold3 inference. Figure~\ref{fig:3}B illustrates these two successful cases, along with examples of backbones generated by partial diffusion that either did not satisfy the mRMSD metric condition or the distinguishability from native. 
%with details of which metric was not satisfied.

\subsection{Example 2: binding site}

Barstar is a small protein that binds the active site of barnase to prevent the latter from breaking RNA. This toxin-antitoxin pair (PDB entry: 7MRX) has been increasingly exploited in cancer therapy for targeted cytotoxicity~\citep{Kalinin2023}. As target for motif diversification, we took the barstar's binding interface region comprising residues 25 to 46 (see~\cite{Watson2023}).
%that diversified barstar's binding interface using PGEL and generated variants with altered affinity. As specified in~\cite{Watson2023}, the motif includes residues 25 to 46.

\begin{figure}[h]
\begin{center}
%\framebox[4.0in]{$\;$}
%\fbox{\rule[-.5cm]{0cm}{4cm} \rule[-.5cm]{4cm}{0cm}}
\includegraphics[width=\linewidth,clip,trim=0cm 7cm 15.5cm 0cm]{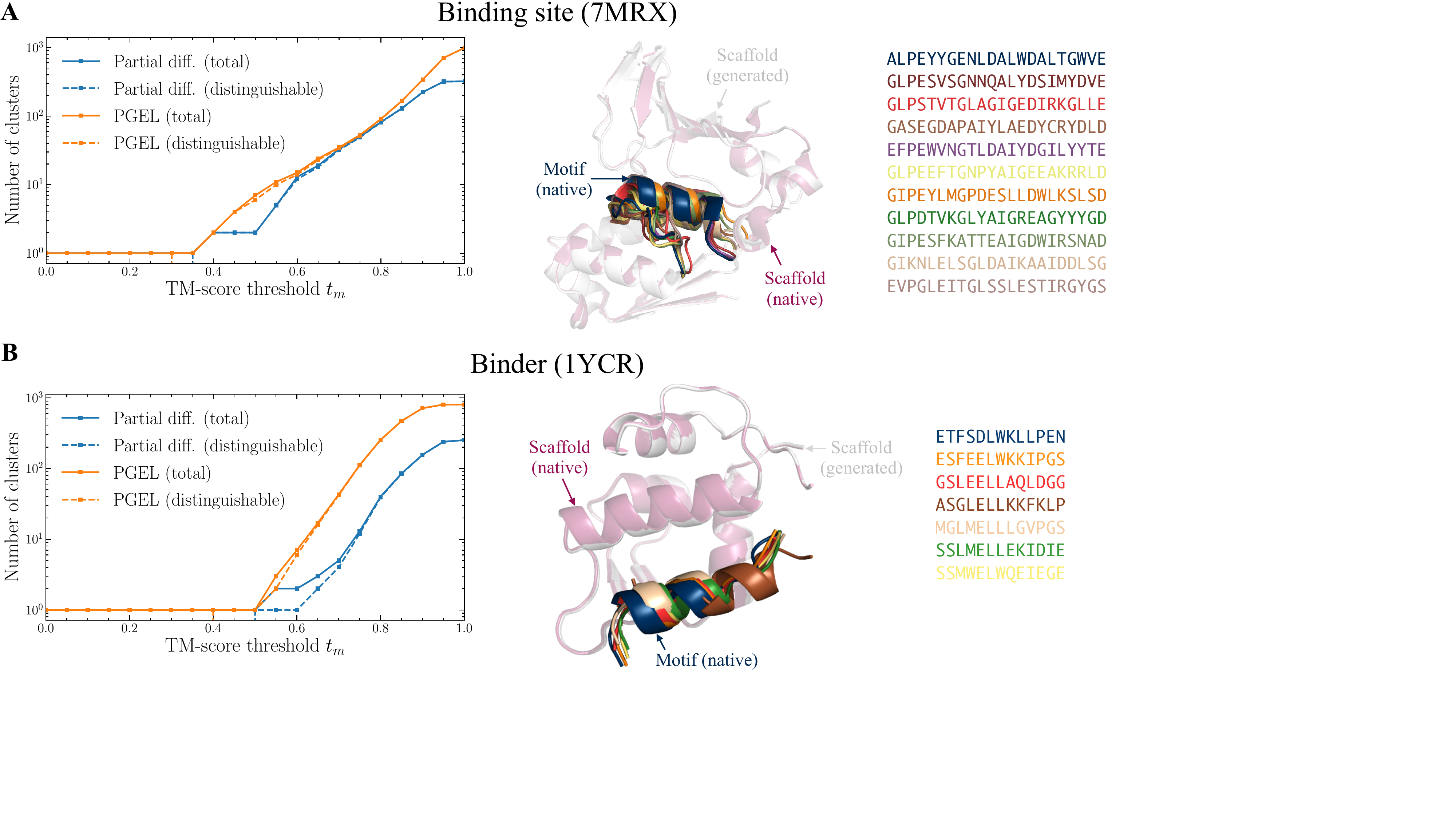}
\end{center}
\caption{(A) Example 2: \textit{Left:} Number of clusters identified PGEL and partial diffusion in the binding site example, both total and distinguishable from native, as a function of the TM-score threshold $t_m$. \textit{Right:} generated binding site backbones (overlaid with native), alongside the native sequence and sequences that refold to self-consistent structures. (B) Same as A but for example 3.} \label{fig:4}
\end{figure}

Of the 1000 backbones generated with PGEL, 990 were classified as designable, nearly tripling the 331 obtained with partial diffusion. Correspondingly, Figure~\ref{fig:4}A demonstrates that PGEL consistently outperforms partial diffusion across the entire range of $t_m\in[0,1]$, with pronounced differences observed at $t_m>0.9$ and within $0.45<t_m<0.55$, around canonical TM-score thresholds. At $t_m=0.6$, PGEL yielded 15 structural clusters compared to 13 for partial diffusion, which were reduced to 10 and 6, respectively, after self-consistency checks (Table~\ref{methods-table}).

In Figure~\ref{fig:4}A, we also display an overlay version of the native motif and 10 representative motifs derived from these clusters, highlighting the sequence variability both among generated barstar binding interfaces and relative to the native PDB structure. When predicting \textit{in silico} the binding affinity of the generated complexes with \textsc{PRODIGY}, two of the generated structures exhibited binding affinities higher than the native complex (see Figure~\ref{fig:5}A and Table~\ref{table:affinity}), while the remaining eight retained at least $80\%$ of the original affinity $\Delta G_{\mathrm{native}}$. In contrast, only one complex generated by partial diffusion had a binding affinity value comparable to that of the native ($\Delta G_{\mathrm{design}}>0.9\Delta G_{\mathrm{native}}$).
%as in that case $\Delta G_{\mathrm{design}}>0.9\Delta G_{\mathrm{native}}$.

\subsection{Example 3: binder}

The interaction between the transcription factor p53 and its negative regulator MDM2 is a key molecular process in cancer progression. Specifically, pharmacological disruption of the p53-MDM2 complex restores p53 activity and has been proven beneficial in cancer therapy~\citep{Hu2021}. 

Starting from PDB entry 1YCR, we addressed the motif diversification task by redesigning the complete p53 under the RMSD constraints described in Section~\ref{subsec:metrics}. PGEL generated 802 designable backbones out of 1000  trials, with most non-designable cases attributable to low pLDDT confidence scores (Table~\ref{methods-table}). Partial diffusion produced only 252 designable backbones with considerably reduced structural diversity (a single cluster at TM-score threshold $t_m=0.6$, see Figure~\ref{fig:4}B). PGEL, by comparison, gave six clusters at $t_m=0.6$, all of which passed the self-consistency checks. 

\begin{figure}[h]
\begin{center}
%\framebox[4.0in]{$\;$}
%\fbox{\rule[-.5cm]{0cm}{4cm} \rule[-.5cm]{4cm}{0cm}}
\includegraphics[width=0.9\linewidth,clip,trim=0cm 26cm 24.75cm 0cm]{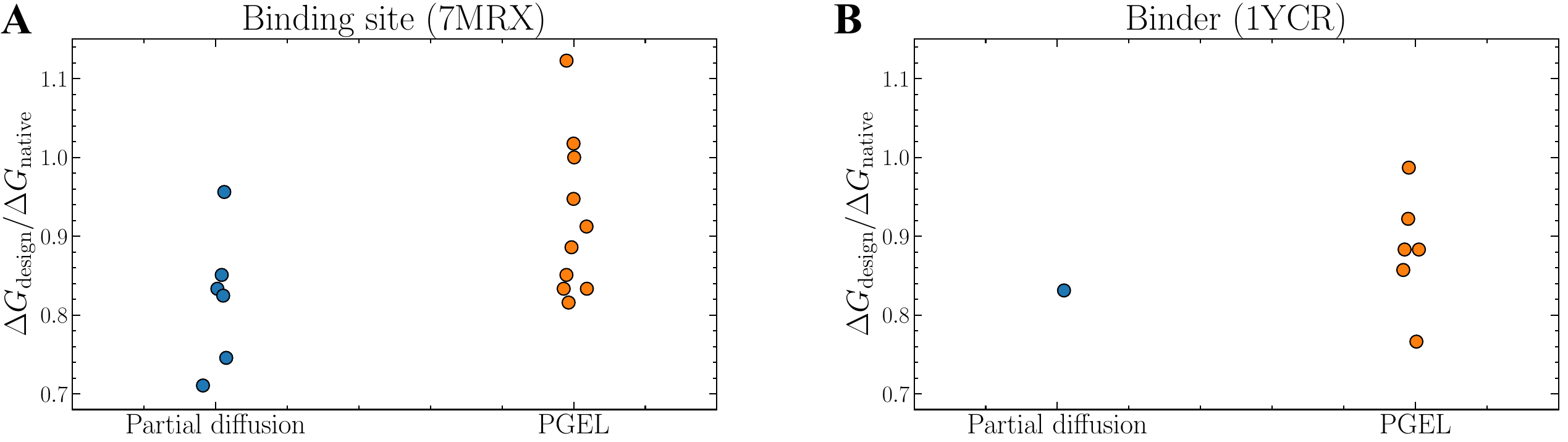}
\end{center}
\caption{Ratios of the binding affinity predicted with \textsc{PRODIGY} of the native structure \textit{vs} the AlphaFold3-predicted structures from backbones (one per cluster) generated by PGEL and partial diffusion for example 2 (A) and example 3 (B).}
\label{fig:5}
\end{figure}

When assessing the binding affinity \textit{a posteriori}, five out of six representatives of PGEL clusters exhibited lower affinity compared to the sole valid instance of partial diffusion (Figure~\ref{fig:5}B, Table~\ref{table:affinity}). Notably, sequence \texttt{SSMWELWQEIEGE} (see Figure~\ref{fig:4}B), designed with PGEL in combination with ProteinMPNN, folded, as predicted by AlphaFold3, into a structure with a binding affinity comparable to that of the native structure, despite sharing only around $15\%$ of sequence identity. This result highlights PGEL's ability to generate backbones that can accommodate sequences unrelated to the native while refolding into structures that preserve function.

\textbf{Computational remarks.} Across the three case studies, column masking was more effective 
%in generating diverse backbones 
than row masking
%.  As shown in 
(see Table~\ref{table:all-results}, with $80\%$ of successful outcomes with $\omega=1$).

The time required per timestep during the generation process is nearly indistinguishable between PGEL and partial diffusion:
%. When running the three examples, the 
average timestep $0.7 \, s$ for partial diffusion and $0.71 \, s$ for PGEL on a single NVIDIA GeForce RTX 3090 GPU with 24GB of memory.

\section{Limitations and future work}

PGEL inherits the biases and limitations of the underlying frozen RFdiffusion denoiser, including its training data distribution and architectural constraints. Moreover, learning embeddings requires additional optimization time, which ranged in our examples from 2 minutes (example 2) to 2 hours (example 3) on a single GPU, with a trade-off between speed and improved results. Our experiments were limited to motifs of up to 40 residues, with practical limits of around 50 residues given available memory, though scaling to longer motifs should be feasible with larger hardware or engineering optimization. Beyond this, our evaluation is entirely \textit{in silico} (pLDDT/RMSD/TM-score filtering, AlphaFold3 refolding, and \textsc{PRODIGY} $\Delta G$) and thus predictive rather than experimental.

Future work will investigate alternative ways of perturbing embeddings, as this strategy for motif diversification remains largely unexplored, as well as different strategies for sampling the masking parameters $\omega$ and $\alpha$. For instance, instead of sampling $\alpha$ uniformly between 0 and 1, one could bias it toward smaller masking rates (\textit{e.g.}, using a Poisson distribution with rate $\lambda$, where $\lambda$ tunes how conservative or aggressive the masking is), thus providing finer control over structural perturbations. 
A more systematic mapping between PGEL's $\omega$ and $\alpha$ and partial diffusion's $T$ would also clarify the relationship between the diversity-fidelity trade-off in both methods. Finally, experimental validation will be pursued in follow-up work.

\textbf{Code availability.} Upon publication, we will release code and configurations to facilitate reproducibility. 

%Please prepare PostScript or PDF files with paper size ``US Letter'', and not, for example, ``A4''. The -t letter option on dvips will produce US Letter files.

%Consider directly generating PDF files using \verb+pdflatex+ (especially if you are a MiKTeX user). PDF figures must be substituted for EPS figures, however.

%Otherwise, please generate your PostScript and PDF files with the following commands: 
%\begin{verbatim}
%dvips mypaper.dvi -t letter -Ppdf -G0 -o mypaper.ps
%ps2pdf mypaper.ps mypaper.pdf
%\end{verbatim}

\subsubsection*{Author Contributions}
Study concept and design: K.M., C.J., A.M.; Development of source
code: K.M.; Analysis and interpretation of data: K.M., A.M.; Writing and revision of the manuscript: K.M., C.J., P.T., T.D., M.B., B.B. and A.M.;
Study supervision: C.J., M.B., B.B. and A.M.

\subsubsection*{Acknowledgments}
K.M. acknowledges support from the President's PhD Scholarship at Imperial College London. M.B. acknowledges support by the Engineering and Physical Sciences Research Council (EPSRC) under grant EP/N014529/1 funding the EPSRC Centre for Mathematics of Precision Healthcare at Imperial College London.
All authors are grateful to Anshul Kanakia, Dino Oglic, Lorena Roldán Martín and Talip Uçar for helpful discussions.

\bibliography{iclr2026_conference}
\bibliographystyle{iclr2026_conference}

\newpage

\appendix
\section{ReverseStep algorithm}\label{appendix:algo}

Let $x^{(t)} = \{(r_l^{(t)}, u_l^{(t)})\}_{l=1}^L$ denote the noisy protein backbone structure at diffusion step $t$, where each residue $l$ is represented by a rotation $r_l^{(t)} \in SO(3)$, with $SO(3)$ the special orthogonal group in three dimensions, and a translation $u_l^{(t)} \in \mathbb{R}^3$. Let $\hat{x}^{(0)} = \{(\hat{r}_l^{(0)}, \hat{u}_l^{(0)})\}_{l=1}^L$ denote the predicted denoised structure. Let $\{\beta^{(t)}\}_{t=1}^T$ be a variance schedule with $\gamma^{(t)} = 1 - \beta^{(t)}$ and $\bar{\gamma}^{(t)} = \prod_{s=1}^t \gamma^{(s)}$. For translations, let %z
$u_l^{(t-1)}$ be sampled from a Gaussian distribution with covariance $\beta^{(t)} I_3$. For rotations, let $s_l$ denote the score approximation presented in \cite{Watson2023}, $\epsilon_{l,d}$ isotropic Gaussian perturbations and $\{f_d\}_{d=1}^3$ a basis of the Lie algebra $SO(3)$. 

\begin{algorithm}[H]
   \caption{\textsc{ReverseStep} function \citep{Watson2023}}
   \label{alg:ReverseStep}
\begin{algorithmic}
   \STATE {\bfseries Input:} noisy structure $x^{(t)}$, denoised prediction $\hat{x}^{(0)}$.
   \STATE {\bfseries Output:} updated structure $x^{(t-1)}$.
   \FOR{$l=1,\dots,L$}
   \STATE $(r_l^{(t)}, u_l^{(t)}) = x_l^{(t)}$
   \STATE $(\hat{r}_l^{(0)}, \hat{u}_l^{(0)}) = \hat{x}_l^{(0)}$
   \STATE $u_l^{(t-1)} \sim \mathcal{N}\!\left(
      \frac{\sqrt{\bar{\gamma}^{(t-1)}}\beta^{(t)}}{1-\bar{\gamma}^{(t)}} \hat{u}_l^{(0)} +
      \frac{\sqrt{\gamma^{(t)}}(1-\bar{\gamma}^{(t-1)})}{1-\bar{\gamma}^{(t)}} u_l^{(t)},\,
      \beta^{(t)} I_3
   \right)$
   \STATE // Updating rotations below
   \STATE $s_l = \textsc{RotationScoreApproximation}(r_l^{(t)}, \hat{r}_l^{(0)}, \sigma_t^2)$
   \STATE $\epsilon_{l,1}, \epsilon_{l,2}, \epsilon_{l,3} \overset{\text{iid}}{\sim} \mathcal{N}(0,1)$
   \STATE $r_l^{(t-1)} = r_l^{(t)} \exp_{I_3}\!\Big\{
      \big(\sigma_t^2 - \sigma_{t-1}^2\big) r_l^{(t)\top} s_l
      + \sqrt{\sigma_t^2 - \sigma_{t-1}^2} \sum_{d=1}^3 \epsilon_{l,d} f_d
   \Big\}$
   \STATE $x_l^{(t-1)} = (r_l^{(t-1)}, u_l^{(t-1)})$
   \ENDFOR
   \STATE \textbf{Return} $x^{(t-1)}$
\end{algorithmic}
\end{algorithm}
\newpage
\section{PGEL evaluation results}\label{appendix:results}

\begin{table}[h!]
\centering
\caption{Detailed results of PGEL successes for examples 1, 2 and 3.}
\begin{tabular}{c c c c c c c c c c}
\toprule
\shortstack{PDB \&\\ design ID} & \raisebox{1.2ex}{$\alpha$} & \raisebox{1.2ex}{$\omega$} & \shortstack{mRMSD \\ (\angstrom)}  & \shortstack{Motif\\pLDDT} & \raisebox{1.2ex}{Sequence} & \shortstack{mRMSD \\ AF3 (\angstrom) } & \raisebox{1.2ex}{pAE} \\
\midrule
\multicolumn{9}{l}{\textbf{1PRW}} \\
17  & 0.75 & 1 & 1.78 & 79 & \resizebox{0.26\textwidth}{!}{FRVIAGGEDGLVTLEQLARY/VRRVAGRGGRLISFEDFLAI} & 1.54 & 4.43 & \\
860 & 0.5 & 1 & 1.96 & 81 & \resizebox{0.26\textwidth}{!}{ARWLDKGGSGAVFGEQLGEF/VAAALEGGKEAKLEEWFLNY} & 1.25 & 4.84 & \\
\midrule
\multicolumn{9}{l}{\textbf{7MRX}} \\
0   & 0.55 & 0 & 0.74 & 79 & \resizebox{0.26\textwidth}{!}{GLPESVSGNNQALYDSIMYDVE} & 0.89 & 3.17 & \\
31  & 0.4 & 0 & 1.35 & 78 & \resizebox{0.26\textwidth}{!}{GLPDTVKGLYAIGREAGYYYGD} & 0.83 & 3.30 & \\
100 & 0.95 & 1 & 1.26 & 73 & \resizebox{0.26\textwidth}{!}{GLPSTVTGLAGIGEDIRKGLLE} & 1.78 & 3.73 & \\
114 & 0.75 & 1 & 0.83 & 75 & \resizebox{0.26\textwidth}{!}{GASEGDAPAIYLAEDYCRYDLD} & 1.22 & 3.79 & \\
145 & 0.4 & 0 & 0.78 & 78 & \resizebox{0.26\textwidth}{!}{EFPEWVNGTLDAIYDGILYYTE} & 0.69 & 4.93 & \\
308 & 0.5 & 1 & 1.57 & 74 & \resizebox{0.26\textwidth}{!}{GIPESFKATTEAIGDWIRSNAD} & 1.25 & 4.97 & \\
352 & 0.85 & 1 & 1.53 & 75 & \resizebox{0.26\textwidth}{!}{GLPEEFTGNPYAIGEEAKRRLD} & 1.98 & 3.81 & \\
730 & 0.8 & 1 & 1.85 & 72 & \resizebox{0.26\textwidth}{!}{GIPEYLMGPDESLLDWLKSLSD} & 1.42 & 4.90 & \\
744 & 0.8 & 1 & 0.88 & 75 & \resizebox{0.26\textwidth}{!}{EVPGLEITGLSSLESTIRGYGS} & 1.96 & 2.42 & \\
814 & 0.3 & 1 & 0.77 & 78 & \resizebox{0.26\textwidth}{!}{GIKNLELSGLDAIKAAIDDLSG} & 1.13 & 3.79 & \\
\midrule
\multicolumn{9}{l}{\textbf{1YCR}} \\
14  & 0.3 & 1 & 1.36 & 75 & \resizebox{0.26\textwidth}{!}{GSLEELLAQLDGG} & 1.56 & 2.88 & \\
36  & 0.7 & 1 & 1.09 & 72 & \resizebox{0.26\textwidth}{!}{MGLMELLLGVPGS} & 1.25 & 3.19 & \\
285 & 0.3 & 1 & 1.59 & 72 & \resizebox{0.26\textwidth}{!}{ASGLELLKKFKLP} & 1.60 & 3.53 & \\
334 & 0.65 & 1 & 1.03 & 72 & \resizebox{0.26\textwidth}{!}{SSLMELLEKIDIE} & 1.25 & 2.88 & \\
619 & 0.2 & 1 & 0.59 & 87 & \resizebox{0.26\textwidth}{!}{ESFEELWKKIPGS} & 1.70 & 2.45 & \\
695 & 0.25 & 1 & 1.04 & 74 & \resizebox{0.26\textwidth}{!}{SSMWELWQEIEGE} & 0.86 & 2.42 & \\
\bottomrule
\end{tabular}
\label{table:all-results}
\end{table}

\newpage
\section{Binding affinity results}

\begin{table}[h!]
\centering
\caption{\textsc{PRODIGY}-predicted binding affinities for examples 2 and 3.}
\begin{tabular}{c c c}
\toprule
\shortstack{PDB \& \\ design ID} & \raisebox{1.2ex}{Method} & \shortstack{\( \Delta G \) \\ (kcal/mol)} \\
\midrule
\multicolumn{3}{l}{\textbf{7MRX}} \\
Native   & -- & -11.4 \\
0   & PGEL & -9.3 \\
31  & PGEL & -11.4 \\
100 & PGEL & -10.8 \\
114 & PGEL & -9.5 \\
145 & PGEL & -10.1 \\
308 & PGEL & -10.4 \\
352 & PGEL & -12.8 \\
730 & PGEL & -11.6 \\
744 & PGEL & -9.5 \\
814 & PGEL & -9.7 \\
1 & Partial diff. & -9.5 \\
18 & Partial diff. & -10.9 \\
307 & Partial diff. & -8.5 \\
327 & Partial diff. & -8.1 \\
513 & Partial diff. & -9.7 \\
780 & Partial diff. & -9.4 \\
\midrule
\multicolumn{3}{l}{\textbf{1YCR}} \\
Native  & -- & -7.7 \\
14  & PGEL & -7.1 \\
36  & PGEL & -5.9 \\
285 & PGEL & -6.8 \\
334 & PGEL & -6.8 \\
619 & PGEL & -6.6 \\
695 & PGEL & -7.6 \\
101 & Partial diff. & -6.4 \\
\bottomrule
\end{tabular}
\label{table:affinity}
\end{table}

\end{document}